\documentclass[
reprint,
 aps,
prb,
]{revtex4-1}

\usepackage{graphicx}
\usepackage{dcolumn}
\usepackage{bm}
\usepackage{amsmath}

\begin{document}

\preprint{APS/123-QED}

\title{Kinetic and thermodynamic temperatures in quantum systems}


\author{Alessio Gagliardi, Alessandro Pecchia$^\dagger$, Aldo Di Carlo}

\affiliation{ Dipartimento di Ingegneria Elettonica, Universita' of Roma "Tor Vergata", Via del Politecnico 1, 00133, Rome, Italy \\
gagliardi@ing.uniroma2.it, Tel: (+39) 06 7259-7367, Fax: (+39) 06 7259-7939. \\
$^\dagger$ CNR-ISMN, Via Salaria km 29.600, 00017 Monterotondo, Roma
}%

\date{\today}

\begin{abstract}
  In this work we present a formalism to describe non equilibrium conditions in systems with a discretized energy spectrum, such as quantum systems.
We develop a formalism based on a combination of Gibbs-Shannon
entropy and information thermodynamics that arrives to a
generalization of the De-Brujin identity applicable to discrete
and non-symmetric distributions. This allows to define the concept
of a thermodynamic temperature with a different, albeit
complementary meaning to the equilibrium kinetic temperature of a
system. The theory is applied to Bosonic and Fermionic cases
represented by an harmonic oscillator and a single energy state,
respectively. We show that the formalism correctly recovers known
results at equilibrium, then we demonstrate an application to a
genuine non equilibrium state: a  coherent quantum oscillator.
\begin{description}
\item[PACS numbers] May be entered using the \verb+\pacs{#1}+
command.
\end{description}
\end{abstract}

\maketitle


Non equilibrium thermodynamics is intimately linked to the concept
of non-equilibrium temperature which in turn tights to the concept
of non-equilibrium entropy. Several attempts have been made in the
past in order to provide a consistent definition of all these
concepts
\cite{tempneq,crooks1,crooks2,jarzynski1,jarzynski2,sagawa1,sagawa2}
and the field is still reach of fruitful debates. One of such
attempts can be called informational thermodynamics, resting on
the extension of the Gibbs-Shannon entropy to non-equilibrium
conditions \cite{jaynes,jaynes2}. This encounters the difficulty
of assigning probabilities to microstates which is solved, for
instance, using the debated postulate of maximal entropy or
maximal entropy production rate. However, for many non-equilibrium
systems, maximal entropy principle fails, signifying that fast
evolving dynamics prevent any meaningful definition of such
probabilities. Nonetheless, in non-equilibrium systems under
steady-state conditions the concept of a probability distribution
is meaningful. Such systems are characterized by steady-state
fluxes of energy, mass or charges driven by non-equilibrium
distributions of phase-space degrees of freedom. To this class
belong a broad range of interesting physical systems ranging from
electronic devices, electrochemical cells, catalytic systems,
photochemical steady state reactions. Especially brought under
attention by the advent of nanotechnology are extremely
miniaturized devices comprising a small number of atoms or
molecules, usually kept under strong non-equilibrium steady-state
conditions. Many examples can be easily found in the literature,
such as nano-mosfets or, at the extreme, molecular electronic
devices \cite{nacho,diventra,nature}. Many assumptions widely used
in the simulation of macroscopic or mesoscopic scales often assume
local quasi-equilibrium,  with a well defined local
electrochemical potential controlling state occupancy and
densities and a local equilibrium temperature \cite{tibercad}.
However, this approximation fails when the perturbation cannot be
considered small with respect to the characteristic length scale.
On the other hand, in a non-equilibrium theory, it would be useful
to retain as much as possible those concepts that have provided so
deep insights in systems close to equilibrium, such as the concept
of temperature, that can be viewed as a parameter controlling
thermodynamics heat fluxes.

A central quantity in statistical mechanics is the Shannon-Gibbs classical entropy, defined as
\begin{equation}\label{ent1}
    S = -k_B \int p(\vec{x}) \log (p(\vec{x})) d\vec{x},
\end{equation}
where $k_B$ is the Boltzmann constant and $p(\vec{x})$ the
probability of the system to occupy the $\vec{x}$ microstate of
the phase space.  The microstate can be expressed in terms of
generalized coordinates representing the degrees of freedom (DOF)
of the system.  Under Gibbs definition, $S$
is a special way of measuring the phase-space available to the
system. We note that equation (\ref{ent1}) is well-defined when the integration is actually interpreted as a dense sum.

In information theory (Shannon entropy), the state $\vec{x}$
represents a set of symbols (word) emitted by an information
source and $k_B=1$. Despite this formal identity, the connection between
thermodynamics and information theory is not obvious but it has
been established \cite{merhav}. Theorems and
formalisms derived in the context of information theory can be
readapted to statistical mechanics.

Entropy and average energy, $\bar{E}$ = $\int
p(\vec{x})\mathcal{E}(\vec{x})d\vec{x}$, define two concepts of temperature. The
thermodynamic temperature is defined as
\begin{equation}\label{thermo}
    T_{th} = \partial \bar{E}/ \partial S,
\end{equation}
and it is related to the statistics in the phase space. The
kinetic temperature, on the other hand, is connected to the energy
stored in the dynamical DOFs. In equilibrium it is directly
connected to the equipartition theorem, giving for a classical gas
of $N$ particles,
\begin{equation}\label{kinetic}
    T_{kin} = \frac{2\bar{E}}{3Nk_B}.
\end{equation}

Finally, for systems in contact with a bath, it is possible to define the temperature of 
the external environment reservoir, $T_0$.


Under equilibrium conditions  $T_{kin}$ = $T_0$ =$T_{th}$. For systems under non-equilibrium these three temperatures assume different values with  different meaning. 
We present here a formalism that extends thermodynamics and temperature to
non equilibrium conditions, at least under steady state, making
extensive use of informational theory concepts. The formalism is
first introduced and applied to an ensemble of quantum harmonic
oscillators following bosonic statistics  (BS) and a single state
ensemble following Fermi statistics (FS), for which the
equilibrium probability distributions (PD) are well known. Then,
we consider the non-equilibrium distribution represented by the
coherent states of a bosonic oscillator, for which we demonstrate
that the thermodynamic, kinetik and bath  temperatures departs
from each others with different meanings
\cite{sanistrava1,alessio1}.

Before moving forward we point out that the quantum extension of (\ref{ent1}) is the Von Neuman entropy based on the density matrix (DM), $S=-k_B \text{Tr}\left[ \hat{\rho}\text{ln} \hat{\rho}\right]$. The theory we discuss here is valid for systems in a mixed state with diagonal DM, reducing the Von Neuman entropy to a classical sum over the probability distribution.
 
An incoherent ensamble of harmonic oscillators in equilibrium follow the geometric distribution,
\begin{equation}\label{bose}
    p_n = e^{-n\frac{\hbar\omega}{k T_0}} \left ( 1 - e^{-\frac{\hbar\omega}{k T_0}}
    \right ) = \Delta^{n} (1 - \Delta).
\end{equation}
Similarly, we can define the equilibrium occupation probability of a single energy state following Fermi statistics,
\begin{equation}\label{p0f}
    p_1 = \frac{1}{e^{\frac{\hbar\omega}{k T_0}} + 1},
\end{equation}
and $p_0 = 1 - p_1$. From these expressions of $p_n$, we can easily verify that  in equilibrium in both cases $T_{th}=T_0$.

In a recent work \cite{alessio1} we have made an extensive
discussion concerning the generalization of entropy for system under steady-state non equilibrium conditions.
In essence, the DOF of a system
driven out of equilibrium correlate to external variables,
$\vec{y}$. As an example we can think of an
external field that induces fluxes of particles or energy,
inducing correlations between particle motions. 
The non-equilibrium form of the energy spectrum and the PD will also be different from equilibrium.
Assuming the from of the Gibbs-Shannon entropy still applicable, non equilibrium (NE) entropy is actually
a conditional entropy,
\begin{equation}\label{eq1}
    S(\vec{x}|\vec{y}) = S^{eq}(\vec{x}) - kI(\vec{x} \wedge \vec{y}),
\end{equation}
where $S^{eq}$ is the equilibrium entropy and the last term is the
mutual information \cite{cover},
\begin{equation}\label{mutual}
    I(\vec{x} \wedge \vec{y}) = \int p(\vec{x},\vec{y}) \log \left ( \frac{p(\vec{x},\vec{y})}{p(\vec{x})p(\vec{y})} \right )
    d\vec{x}d\vec{y}.
\end{equation}
Equation (\ref{eq1}) shows that a system out of equilibrium can be
mapped into a system in equilibrium with additional information
\cite{alessio1}. The same equation also implies a reduction of the
available phase space, associated with a reduction of entropy and
increase of available free energy, according to:
\begin{equation}\label{fne}
    F_{NE} = \bar{E}_{NE} - T_{th}S^{eq} + k_B T_{th} I,
\end{equation}
where $\bar{E}_{NE}$ is the average non-equilibrium energy
(depending on the NE PD). 

We turn now our attention to discrete PDs, $p_n$, defined on a
generally finite support, $n\in\left[ \alpha, \beta \right]$, that
may represent a discrete energy spectrum. For discrete
distributions the integrals of equations (\ref{ent1}) and
(\ref{mutual})  become summation over the index $n$.
In order to evaluate (\label{thermo}) we need to construct a functional perturbation to 
 the PD, $p_n$, that induces an infinitesimal increase of entropy, $\delta S$, and mean energy, $\delta
E$. In making such a mathematical construction we assume that the energies of the microstates, $E_n$,
are left unchanged. This is consistent with taking the derivative
of eq. (\ref{thermo}) at constant volume and particle number.
In a way that resembles the original De Brujin scheme for continuous PD, we consider a perturbation obtained as a discrete convolution $p^{\varepsilon}_n=\sum_m p_{n+m}q_m(n,\varepsilon)$, where
$q_m(\varepsilon)$ is generally defined in the interval $\alpha
<n< \beta$ as \cite{sanistrava2},
\begin{equation}\label{p2}
\begin{cases}
 q_{-1}= \varepsilon  \\
 q_{0}=   1 - 2\varepsilon \\
 q_{+1}= \varepsilon
\end{cases}
\end{equation}
and $q_{m}=0$ for $m>1$ and $m<-1$.  However the perturbation needs to be modified at the two boundaries such
that for $n =\alpha$, $q_{-1}= 0$, $q_{0} = 1 - \varepsilon$, $q_{1} = \varepsilon$ and for $n =\beta$, $q_{-1} = \varepsilon$, $q_0 = 1 - \varepsilon$, $q_1 = 0$.
It follows that the perturbed PD, $p^{\varepsilon}$, is
\begin{equation}
\begin{cases}
   p^{\varepsilon}_{\alpha} = p_{\alpha} + \varepsilon ( p_{\alpha+1} - p_{\alpha}) \\
   p^{\varepsilon}_{n} = p_n + \varepsilon (p_{n+1} -2p_{n} + p_{n-1} ), &  \alpha <n< \beta \\
   p^{\varepsilon}_{\beta} = p_{\beta} + \varepsilon ( p_{\beta-1} - p_{\beta} ).
\label{pert1}
\end{cases}
\end{equation}
The modifications at the boundaries are just
needed for preserving the normalization of the PD to unity. For a Fermi statistics the perturbed probability reduces to
$p^{\varepsilon}_{0} = (1 - \varepsilon)p_{0} + \varepsilon p_{1}$ and
$p^{\varepsilon}_{1} = (1 - \varepsilon)p_{1} + \varepsilon p_{0}$.
We note that the original De Brujin identity is demonstrated for continuous PDs defined on a infinite support. In this contex $q_m$ becomes a normal distribution with mean 0 and variance $\varepsilon$.    

{\bf Theorem 1:} Given a perturbation of PD as in equations (\ref{pert1}), in the limit $\varepsilon \rightarrow 0^{+}$,  the entropy variation can be written as
\begin{equation}\label{temp234}
  \frac{1}{k}   \frac{\delta S}{\varepsilon}  =  D \left[ p_{n+1}\|p_{n} \right] + D \left[ p_{n-1}\|p_{n} \right],
\end{equation}
where $D\left[p_{n+1}\|p_n\right]$ is the Kullblak-Leibler
divergence (KLD), a well-known concept in information theory
\cite{alessio2}, defined as
\begin{equation}\label{kl}
    D\left[p_{n}\|q_n\right] = \sum_n p_n \ln \left( \frac{p_n}{q_n} \right),
\end{equation}
representing a pseudo-distance between two PDs, which
has the property of being always non negative and equal to zero iif $p_n$ = $q_n$,  $\forall n$. The demonstration of this theorem is given in the appendix.

Observing that $D\left[p_{n}\|p_{n}\right]=0$, the r.h.s. of
equation (\ref{temp234}) can be rewritten as $D \left[
p_{n+1}\|p_{n} \right] + D \left[ p_{n-1}\|p_{n}\right] -2
D\left[p_{n}\|p_{n}\right]$, which has the suggestive form of a
discrete curvature measure of the PD. This geometrical concept has
interesting implications developed by Amari and coworkers
\cite{amari}. The perturbed PD also gives a perturbation to the
average energy according to
\begin{eqnarray}
    \frac{\delta \bar{E}}{\varepsilon} & = & \sum_{n} E_{n}\frac{p^{\varepsilon}_{n} -p_{n}}{\varepsilon}.
\label{ene12}
\end{eqnarray}
Combining eq. (\ref{temp234}) and (\ref{ene12}) we can rewrite
$\frac{dS}{d\bar{E}}=\frac{\delta S}{\varepsilon}
\frac{\varepsilon}{\delta \bar{E}}$ giving,
\begin{equation}\label{temp123}
\frac{1}{kT_{th}} \frac{\delta \bar{E}}{\varepsilon} = D\left[p_{n+1}\|p_{n}\right] + D\left[ p_{n-1}\|p_{n}\right].
\end{equation}
A similar expression was obtained  in \cite{sanistrava2,prl}, but
in those derivations the PD is always assumed defined on an
infinite support $n\in(-\infty,\infty)$ and completely symmetric
with respect to the origin. Under the more restrictive assumptions
stated above, equation (\ref{temp123}) leads to the well-known De
Brujin identity \cite{cover}, usually derived in the context of
continuous and differentiable PDs. In such a context $T_{th}$
is also known as the Fisher temperature of the system. 
Our theorem can be considered a generalization to non
symmetric and discrete PDs.
The generality of {\em Theorem 1} makes it valid to any arbitrary
PD, including non equilibrium conditions. Indeed, no specific
assumptions have been made. In the case of distributions defined on semi-infinite supports, i.e. $n\in[0,+\inf)$, 
the form (\ref{p2}) need to be corrected only at the lowest boundary. 

An obvious sanity check is to show that under an equilibrium PD
the relationship (\ref{temp123}) leads to the correct equilibrium
temperature. This can be shown for the Bose-Einstein and Fermi
statistics using the detailed-balance relationships, $p_{n+1} =
p_{n} \Delta$. Substituting in eq. (\ref{ene12}) we get $\delta
\bar{E}/\varepsilon = \hbar\omega( 1 - \Delta )$ from which
follows, from eq. (\ref{temp123}), $ \hbar\omega( 1 - \Delta )
/T_{th}= \hbar\omega ( 1 - \Delta )/T_{0}$, therefore
$T_{th}=T_{kin}=T_0$. A similar proof can also be carried out for
the equilibrium Fermionic PD of eq. (\ref{p0f}).

Under non equilibrium conditions the three temperatures ($T_0$,
$T_{th}$ and $T_{kin}$) are generally different. The kinetic
temperature is in fact obtained from $\bar{E}_{NE}$ alone, while
the thermodynamic temperature is related to the statistics of the
PD. The meaning of these temperatures is still debated but could be generally linked to the heat and mechanical work exchanged between reservoir and system \cite{sanistrava1}. Limited to the assumption of Langivin dynamics, Sanistrava and coworkers demonstrate the relationships,   
\begin{eqnarray}\label{heat}
    \Delta Q \propto T_{0} - T_{th}, \\
    \Delta W \propto T_{th} - T_{kin},
\end{eqnarray}
where $\Delta Q$ and $\Delta W$ are, respectively, the net heat and
mechanical work exchanged. A positive sign corresponds to a system absorbing energy (heat or work) from the reservoir.
The relevant result of the relations above is that $T_{th}$ and $T_{kin}$ could be used to understand fluxes of work and heat in non-equilibrium systems, simply based on the non-equilibrium PD.

{\em Laser Field.} As an example of our formalism we consider the non-equilibrium $T_{th}$ and $T_{kin}$ for a cavity laser.
This can be described as a mixed state given by a superposition of coherent states with a distribution of all possible phases,
\begin{equation}\label{laserdm}
\hat{\rho}=\frac{1}{2\pi}\int d\phi |\alpha e^{i\phi}\rangle\langle\alpha e^{i\phi}|=\sum_n e^{|\alpha|^2}\frac{|\alpha|^{2n}}{n!} |n \rangle\langle n|,
\end{equation}
in which the coherent or Glauber states are defined as
\begin{equation}\label{coherent1}
    |\alpha\rangle = e^{-\frac{|\alpha|^{2}}{2}}\sum_{n} \frac{\alpha^{n}}{\sqrt{n!}}|n\rangle.
\end{equation}
From (\ref{laserdm}) follows that the non-equilibrium laser field can be described by an ensamble following a Poissonian distribution, 
\begin{equation}\label{poissoncoherent}
   r_n = \frac{|\alpha|^{2n}}{n!}e^{-|\alpha|^{2}},
\end{equation}
from which the averaged occupation number is $\langle n \rangle = |\alpha|^{2}$
and the averaged energy is $\bar{E} = \hbar\omega \left( |\alpha|^{2} + \frac{1}{2} \right )$.
Due to this particular relation, the kinetic temperature can be obtained by inverting the equilibrium Bose-Einstein distrubution, 
\begin{equation}\label{kin1}
    T_{kin} = \frac{\hbar\omega}{k_B \ln \left( 1 + \frac{1}{ |\alpha|^{2}}\right)}.
\end{equation}
\begin{figure}[tb]
\begin{center}
\includegraphics*[width=6.5cm, angle=270]{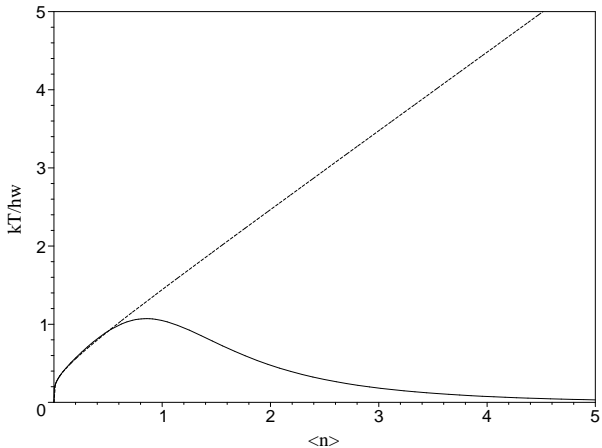}
\caption{Kinetic (dashed) and thermodynamic (solid)
temperatures for the coherent state as a function of the average
occupation number ($kT_0$=26 meV).} \label{fig:temp}
\end{center}
\end{figure}

\begin{figure}[tb]
\begin{center}
\includegraphics*[width=6.5cm, angle=270]{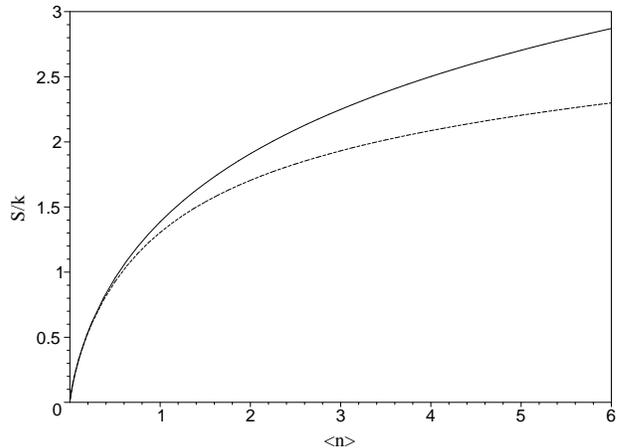}
\caption{ Plot of the equilibrium (solid) and non-equilibrium (dashed) entropies using
eq. \ref{ent1} for the geometric and Poissonian probability distributions ($\hbar \omega$ =1).} \label{fig:entropy}
\end{center}
\end{figure}

Applying eq. (\ref{temp234}) to the Poissonian distribution it is possible to compute
 the thermodynamic temperature for the cavity field. The calculation of the KLD divergences is made simple by using the relationship  $(n+1) r_{n+1} = |\alpha|^{2} r_{n}$, that
after some manipulations, leads to
\begin{equation}\label{temppoisson}
    T_{th} = \frac{\hbar\omega r_{0}}{k \left [ \sum_{k=1}^{\infty} r_{k} \ln \left (\frac{k+1}{k} \right ) - r_0 \ln(|\alpha|^{2}) \right ] },
\end{equation}
with $r_0 =\exp(-|\alpha|^{2})$. There is no closed analytical
form for this $T_{th}$, however the series in the denominator
converges very quickly and can be numerically evaluated.
The thermodynamic and kinetic temperatures for different $\langle n
\rangle$ are shown in Figure \ref{fig:temp} where we can observe that for low values of
$\langle n \rangle$ the two temperatures are essentially equivalent, but then
the kinetic temperature increases with the average energy, while
the thermodynamic temperature goes asymptotically to zero.
This shows that the intuitive concept of temperature is related to $T_{kin}$, whereas $T_{th}$ measures in fact
something quite different, related to the statistics of the PD.
 For a laser field, even if the average energy is increasing with $\langle n \rangle$, the
'disorder' associated with the distribution decreases and with it the thermodynamic temperature. 
As shown in  Figure \ref{fig:entropy}, the
non-equilibrium entropy is strictly lower than the entropy in
equilibrium, as expected from the relationships
(\ref{eq1}-\ref{fne}), for the same average energy or average
occupation,  $ \langle n \rangle $. We also notice that the mutual
information, $k_B I = S^{eq}-S$, increases with  $ \langle n \rangle
$, reflecting the fact that the system free energy increases.
Indeed, manipulating equation (\ref{fne}) we get
$F-F_{eq}=(T_0-T_{th})S^{eq} + k_B T_{th}I$. The fact that
this difference increases with $\langle n \rangle$  reflects the
high degree of order of the energy stored in a coherent state as
opposed to a thermalized equilibrium energy.  From this follows
that, without external constrains, the coherent state will evolve
towards an equilibrium state of maximal entropy.
Concluding, we have developed a formalism to treat non-equilibrium
conditions for discrete distributions  based on a new
generalized relation between the Kullback-Leibler curvature and the thermodynamic temperature. The
formalism has been applied to the equilibrium case of Bose and Fermi distributions to show its
validity and then to a real non equilibrium case represented by a quantum
harmonic oscillator in a coherent state.
The generality of our approach paves the way to a new simple approach to compute non-equilibrium thermodynamical properties,
 that can be applied to more complex systems and nanoscale quantum devices under non-equilibrium conditions.

\section{Appendix A: Theorem demonstration}

We want to compute the entropy variation of equation (\ref{temp234}),
\begin{equation}\label{eqA1}
    \frac{\delta S}{\varepsilon} = \frac{S(p^{\varepsilon}) -
    S(p)}{\varepsilon},
\end{equation}
where the difference is respect the entropy computed with the
perturbed probability distribution $p^{\varepsilon}$ of equation
(\ref{pert1}), and the reference entropy. In 
appendix of ref. \cite{sanistrava2} a demonstration of equation (\ref{temp234}) is
presented, but the reference PD is assumed symmetric and with
infinite support. We extend their results for a general discrete non-symmetric
PD with arbitrary support.


Using the perturbation form of equation (\ref{pert1}), we can write
\begin{eqnarray}
    S[p^{\varepsilon}] & = & -k_B \sum_{n=\alpha + 1}^{\beta -1} [(1 - 2\varepsilon)p_{n} + \varepsilon p_{n+1} + \varepsilon
    p_{n-1}] \ln(p^{\varepsilon}_{n}) \nonumber \\
    & & -k_B [(1 - \varepsilon)p_{\alpha} + \varepsilon
    p_{\alpha+1}]\ln(p^{\varepsilon}_{\alpha}) \nonumber \\
    & & -k_B [(1 - \varepsilon)p_{\beta} +\varepsilon p_{\beta -1}] \ln(p^{\varepsilon}_{\beta}).
\label{pertent1}
\end{eqnarray}
From this follows, rearranging terms,
\begin{eqnarray}
   \delta S & = & k_B D[p_{n}\|p^{\varepsilon}_{n}]  + k_B \varepsilon  \sum_{n=\alpha + 1}^{\beta} (p_{n}-p_{n-1}) \ln(p_n) \nonumber \\
                 &    & + k_B \varepsilon \sum_{n=\alpha}^{\beta -1}  (p_{n}-p_{n+1}) \ln(p_n)
\label{main}
\end{eqnarray}
where we have used the discrete form of the KLD and that $p^{\varepsilon}_{n} = p_{n}+O(\varepsilon)$. The first term can be expanded in series as
\begin{equation}\label{KL2}
  D[p\|p^{\varepsilon}] = D[p\|p] + \frac{\partial D[p\|p^{\varepsilon}]}{\partial \varepsilon}\varepsilon +O(\epsilon^2),
\end{equation}
where  $D[p\|p]=0$ by definition. Thanks to the normalization of
$p_n^{\varepsilon}$ the term $\partial
D[p\|p^{\varepsilon}]/\partial \varepsilon = 0$ in the limit
$\varepsilon \rightarrow 0^{+}$. It follows that
$D[p\|p^{\varepsilon}]$ = $o(\varepsilon^2)$ \cite{sanistrava2}.
The following two terms can be expressed once more using the KLD,
as
\begin{equation}\label{ent23}
   \lim_{\varepsilon \rightarrow 0^{+}} \frac{\delta
   S}{\varepsilon} = k_B D[p_{n+1}\|p_{n}] + k_B D[p_{n-1}\|p_{n}],
\end{equation}
with the understanding
\begin{equation}\label{KL23}
    D(p_{n+1}\|p_n) = \sum_{n=\alpha}^{\beta-1} p_{n+1} \ln \left( \frac{p_{n+1}}{p_{n}}\right).
\end{equation}
\begin{equation}\label{KL23}
    D(p_{n-1}\|p_n) = \sum_{n=\alpha+1}^{\beta} p_{n-1} \ln \left( \frac{p_{n-1}}{p_{n}}\right).
\end{equation}
The last results demonstrate the theorem.




\end{document}